\documentclass[12pt]{article}
\usepackage[subpreambles=true]{standalone}
\usepackage{graphicx}
\usepackage{setspace}
\usepackage[margin=2cm]{geometry}
\usepackage{amsmath, amssymb}
\usepackage{kpfonts}
\usepackage{float}
\usepackage{booktabs}
\usepackage{caption}
\usepackage{xurl}
\usepackage{enumerate}
\usepackage[backend=biber,
            doi=false,
            url=false, 
            isbn=false, 
            eprint=false,
            style=authoryear,
            maxbibnames=99,
            maxcitenames=99,
            ]{biblatex}
\usepackage{hyperref}
\hypersetup{colorlinks}

\title{Rising Marginal Costs, Rising Prices?}

\author{
Joel Kariel\thanks{Competition and Markets Authority \& University of Kent, joel.kariel@cma.gov.uk}
\and
Anthony Savagar\thanks{University of Kent, asavagar@gmail.com.}
}      

\date{30 December 2024}	

\begin{document}
\begin{titlepage}
\clearpage
\maketitle
\thispagestyle{empty}
\begin{abstract}
\onehalfspacing
\noindent 
We present empirical evidence on the relationship between demand shocks and price changes, conditional on returns to scale. We find that in industries with decreasing returns to scale, demand increases (which raise costs) correspond to price increases. Whereas, in industries with increasing returns to scale, demand increases (which lower costs) correspond to stable prices. We interpret the results with a theory of imperfect competition and returns to scale. For prices to remain stable following a cost decrease, markups necessarily rise. For prices to increase as cost increases, it is not necessary for markups to change but does not preclude their role. From a macroeconomic perspective, our results imply that inflation dynamics and the effectiveness of monetary policy depend on market structures.
\\ [1em]
\noindent \textbf{JEL}: E3, L1. 
\\
\textbf{Keywords}: Markups, Marginal Costs, Returns to Scale, Market Structures, Inflation.
\end{abstract}
\end{titlepage}

\onehalfspacing

\section{Introduction}

Understanding how prices respond to demand shocks is crucial for macroeconomic modeling and policy analysis.  While recent literature and policy attention has focused on the role of markups in determining aggregate price behaviour -- often dubbed `greedflation'-- less focus has been given to firm cost schedules, which are determined by returns to scale. 

We investigate the following research question: \textit{How do returns to scale affect the transmission of demand shocks to prices?}
Using UK production data across all sectors of the economy, we show that: in industries with decreasing returns, positive demand shocks correlate with price increases. Whereas, in industries with increasing returns, positive demand shocks do not correlate with prices. This implies that as output expands, and costs increase, due to decreasing returns, the costs are passed through to prices. But, when output expands, and costs fall, due to increasing returns, the cost savings are not passed through to prices. 

Our empirical analysis has two components. First, we estimate returns to scale for ten 1-digit industries based on 3-digit industry observations, and we classify them as increasing, decreasing or constant returns. 
We estimate production functions on industry-level data using several estimation methodologies (OLS, IV, control function). The data for our production function estimation is from the UK's Annual Business Survey (ABS), which is an annual survey of production across all sectors.
Second, we explore the relationship between price and demand changes for industries with increasing and decreasing returns. To do this, we regress price changes on output changes at the 2-digit level separately for increasing and decreasing returns industries. To measure price changes we use 2-digit industry price deflators from the Office for National Statistics (ONS) and for demand changes we use sales changes from the ABS. 

These results form three contributions. First, we emphasize that cost structure affects the transmission of demand shocks to prices, and should be considered alongside competition-based arguments. Second, our results have indirect implications for markup behaviour. With increasing returns, markups do not necessarily correlate with prices, but with decreasing returns markups must correlate positively with prices. 
Hence, if we bundle increasing and decreasing returns industries together, we may observe no relationship on average. Whereas, if an economy consists of decreasing returns industries, markup and price changes will correlate, or if demand shocks are heterogeneous and affect increasing or decreasing returns industries separately, then prices will increase or be stable accordingly. Finally, we use data that is representative (across sector, employment size and geography) of firm population in the entire macroeconomy, rather than being subject to the biases of proprietary datasets that are often used to analyse across many indutries.

\subsubsection*{Related Literature}

\textcite{ConlonMillerOtgonYao2023_AEApp} investigate the relationship between price changes and markup changes across industries. They find that markup changes are not correlated with price changes. This implies that changes in markups are potentially driven by supply-side factors affecting marginal costs, rather than competition factors affecting prices. This analysis and similar suggestions by \textcite{Syverson2019_JEP} motivate our investigation into marginal costs.

Several recent policy papers analyse the effect of energy cost shocks on inflation. The link to our paper is that a demand shock in an increasing returns sector acts as a negative cost shock, while in a decreasing returns sector, it translates into a positive cost shock. \textcite{KharroubiSpigtIganTakahashiZakrajsek2023_BIS} find that firms pass-on positive cost shocks (oil price increases), but do not pass on negative cost shocks, which is consistent with our finding. 
\textcite{ManuelPitonYotzov2024_EL} find that recent energy cost increases in the UK are associated with lower or stable margins, but high-markup firms are insulated. This implies that cost increases are the main determinant of price increases.

Recognition of the role of returns to scale is growing in macroeconomics. Typically, macroeconomic models assume constant returns to scale. This eases aggregation and was supported by empirical evidence on the US following \textcite{BasuFernald1996_JPE}. Recently, several papers have explored the role of returns to scale for aggregate outcomes mostly focusing on real effects on productivity or output in firm dynamics models \parencite{KehrigGao2021_wp, RuzicHo2021_ReStat, Smirnyagin2023_JME, KarielSavagar2024_WP}, but without our focus on price setting. 
Finally, \textcite{AhlanderKleinPappa2024_WP} present new empirical evidence from PPI data which shows that supply curves can be flat or downward sloping, suggesting increasing returns to scale.

\section{Theory}

We divide this section into two parts reflecting textbook producer theory that divides the firm problem into cost minimization and profit maximization. First, firms choose inputs to minimize cost for any given amount of output. Second, firms choose output to maximize profits. The cost minimization stage is independent of demand conditions. It only depends on the properties of the production function. We focus on this stage which is relevant for returns to scale. The profit maximization stage depends on the demand conditions faced by the firm. This is relevant for markup setting behaviour which will differ according to the demand system. Various different demand systems have been implemented in macroeconomic models such as CES (Dixit-Stglitz), oligopolistic competition, Kimball aggregator, translog (Feenstra) preferences. Our empirical results indirectly imply markup behaviour, which may support some theories over others, but we do not attempt to model this.

\subsection{Cost Minimization: Returns to Scale}
Returns to scale are a technical property of the production function. They are not determined by economic decisions or market conditions. They determine the slope of the marginal cost curve which is determined from firms' cost-minimising behaviour. 

Consider a Cobb-Douglas production function, where $\jmath$ indexes an industry. Output $Y_\jmath$ is produced by capital $K_\jmath$, labour $L_\jmath$ and materials $M_\jmath$, combined with an exogenous technology term $A$:
\begin{equation*}
    Y_\jmath = A K_\jmath^\alpha L_\jmath^\beta M_\jmath^\gamma
\end{equation*}
The function is homogeneous of degree $\nu
\equiv \alpha + \beta + \gamma$ in inputs of capital, labour and materials. The parameter $\nu$ captures returns to scale. That is, if inputs increase by a constant $c$, then output will increase by a factor $c^\nu$. Decreasing returns occur when $\nu \in (0,1)$, constant returns occur when $\nu=1$, and increasing returns occur when $\nu \in (1,\infty)$.
The cost function dual to the production function is:
\[\mathcal{C}_\jmath = z  r^\frac{{\alpha}}{\nu} w^\frac{\beta}{\nu} q^{\frac{\gamma}{\nu} }Y_\jmath^{\frac{1}{\nu}},\]
where $z \equiv \nu \left[A \alpha^{\alpha} \beta^\beta \gamma^\gamma \right]^{-\frac{1}{\nu}}$ is a constant. Input prices are exogenous and the same for all agents since there is no factor market power: $w$ is the price of labour, $r$ is the price of capital and $q$ is the price of materials.\footnote{The cost function follows from solving the firm's cost minimization problem $\mathcal{C}_\jmath = \min_{L_\jmath , K_\jmath, M_\jmath} \quad  w L_\jmath + r K_\jmath + q M_\jmath$ subject to $Y_\jmath \leq A K_\jmath^\alpha L_\jmath^\beta M_\jmath^\gamma$. Then, rearranging the optimality conditions as derived demand functions for capital, labour and materials. The derived demands are functions of output, production function parameters and input prices. The cost function follows from substituting the derived demands into the cost definition.} Marginal cost is given by the first-derivative of the cost curve and the second-derivative is the slope of the marginal cost curve:
\[mc_\jmath \equiv \frac{d \mathcal{C}_\jmath}{d y_\jmath} = \frac{1}{\nu} \frac{\mathcal{C}_\jmath}{y_\jmath} \quad \text{and} \quad \frac{d^2\mathcal{C}_\jmath}{dy_\jmath^2} = \frac{1-\nu}{\nu}.\]
The slope of the marginal cost curve is constant, and depends on whether there are constant $\nu=1$, increasing $\nu \in (1,\infty)$ or decreasing $\nu \in (0,1)$ returns to scale

Because there are no fixed costs, returns to scale of the production function are equivalent to the average cost ($\mathcal{C}_\jmath/y_\jmath$) to marginal cost ratio (i.e., the inverse scale elasticity):
\[\left( \frac{d \mathcal{C}_\jmath}{d y_\jmath} \frac{y_\jmath}{\mathcal{C}_\jmath} \right)^{-1} = \nu.\]
The average cost to marginal cost ratio is commonly used to measure scale economies in applied industrial organization \parencite{DavisGarces2009_book}.

\subsection{Profit Maximization: Price, Marginal Cost, Markup Relationship}
Profit maximizing firms produce output where marginal revenue equals to marginal cost. In the case of perfect competition, the marginal revenue is the price, leading price to equal marginal cost. With imperfect competition, price diverges from marginal revenue leading to a wedge between price and marginal cost. This wedge is the price markup, and its properties will depend on the demand system.\footnote{For example, the CES demand system is common in macroeconomic models, and has constant markups $\hat{\mu}=0$ such that price and marginal cost movements are equal $\hat{p}=\hat{mc}$.} Specifically, the price markup $\mu_\jmath$ is the ratio of its price $p_\jmath$ to its marginal cost $mc_\jmath$:
\[\mu_\jmath \equiv \frac{p_\jmath}{mc_\jmath}.\]
A first-order approximation of the markup relationship implies:
\[\hat{p}_\jmath = \hat{\mu}_\jmath + \hat{mc}_\jmath,\]
where hat notation represents differences or percentage deviation from a reference point. Therefore, prices increase with markup increases and marginal cost increases. 

Our main focus is the relationship between price and marginal cost. Following a demand shock, marginal costs move in opposite directions for increasing and decreasing returns industries. With increasing returns, output expansion decreases marginal costs, and with decreasing returns, it increases marginal costs. Therefore, if markup movements are small (weaker than marginal cost movements), we hypothesise:
\begin{enumerate}[(i)]
    \item In increasing returns to scale industries, positive demand shocks decrease price.
    \item In decreasing returns to scale industries, positive demand shocks increase price.
\end{enumerate}
If these relationships hold, it suggests that cost movements are influential for price movements, whilst markup movements are not strictly necessary. Specifically, these relationships rule out large markup movements in the opposing direction to costs. However, they cannot rule out constant or small opposing markup movements, and they cannot rule-out markups reinforcing cost movements.\footnote{A possible `false negative' is that when markups reinforce marginal costs, the positive relationship between price and marginal cost is driven by omitted markups, rather than marginal costs.} However, if the relationships do not hold, such that prices are unresponsive or move in the opposite direction, then markups play a role. Specifically, markups move in the opposite direction to costs and offset or dominate the cost movements.

\section{Data}
To estimate returns to scale, we use sectoral data from the Annual Business Survey (ABS) Standard Extracts 2019. We include data details in the appendix. The data is at the industry level for various SIC levels. The industry data is aggregated from firm responses to the UK's production survey. To construct the aggregate figures the ONS use weights to make the subset of surveyed firms representative of the entire industry. The main variables of interest are gross output, employment, capital and materials. We employ the following data cleaning:
\begin{enumerate}
    \item The ABS data is obtained at the 3-digit by year level, from 2008 - 2019.
    \item Sales, intermediate inputs and net capital are deflated with price, material and capital deflators respectively. These are 2/3-digit by year deflators, obtained from ONS experimental industry deflators,  ONS supply and use tables and ONS capital stocks.
    \item We take logs of deflated sales, deflated intermediate inputs, labour input (number of employees) and deflated net capital. 
    \item Missing employment observations are interpolated by gross value-added (466 of 2,868). 
\end{enumerate}

\section{Empirical Methodology}

We estimate the following regression which corresponds to the Cobb-Douglas production function outlined above:
\[ \ln Y_{\jmath t} = \beta_L \ln L_{\jmath t} + \beta_k \ln K_{\jmath t} + \beta_M\ln  M_{\jmath t} + \epsilon_{\jmath t} \]
To obtain returns to scale estimates, we sum the estimated coefficients, which represent output elasticities. We use the variance-covariance matrix to obtain the standard errors of returns to scale since they are a linear combination of the coefficients.

Our baseline regression uses OLS to estimate the elasticities. In the appendix we provide results using IV \parencite{Hall1988_JPE} and control functions \parencite{LevinsohnPetrin2003_ReStud}. They produce qualitatively similar results. There is an extensive literature on the estimation of production functions, which addresses the endogeneity problem that unobserved productivity affects input choice. Common methods are dynamic panel and control function approaches  \parencite{BlundellBond2000_ER, AckerbergCavesFrazer2015_ecta, GandhiNavarroRivers2020_JPE}.

Finally, we regress annual changes in price deflators on annual changes in sales at the 2-digit industry level. This level of aggregation is selected due to the level of aggregation of price deflators. All differenced variables are standardised (i.e. annual 2-digit sectoral differences are relative to the sector-specific average difference, divided by the sector standard deviation). This regression interacts changes in sales with an indicator for whether the 2-digit sector is in an increasing returns to scale 1-digit industry. For robustness, we include alternative results with year fixed effects and observations weighted by gross value-added or the number of firms.

\section{Empirical Results}
Table~\ref{tab:ols_abs_sectoral} presents our estimates of output elasticities and returns to scale at the 1-digit industry level for 10 broad industries. Each 1-digit industry contains $N$ observations at the 3-digit industry level for a maximum of $T=12$ time periods. Each column of Table~\ref{tab:ols_abs_sectoral} contains output elasticities at the 1-digit SIC level. 
\begin{itemize}
\item The following industries have increasing returns: 
\begin{itemize}
    \item 1: Agriculture, Forestry, Fishing, Mining and Quarrying
    \item 3: Manufacturing (Chemicals, Metals, Machinery)
    \item 7: Transportation, Storage and Communication
    \item 8: Accommodation and Food Service
\end{itemize}
\item The following industries are decreasing
\begin{itemize}
    \item 2: Manufacturing (Food, Beverages, Textiles)
    \item 4: Utilities (Electricity, Gas, Water)
    \item 5: Construction
    \item 6: Wholesale and Retail Trade
    \item 9: Financial, Insurance and Real Estate
    \item 10: Public Administration, Education, Health, Arts
\end{itemize}
\end{itemize}
(3),(4),(5),(8) are not different to constant returns with 95\% statistical significance, though (5),(8) would reject constant returns for slightly weaker statistical significance. (3),(4) fail to reject the constant returns hypothesis even for low levels of significance.
{
\def\sym#1{\ifmmode^{#1}\else\(^{#1}\)\fi}
\begin{table}[H]
\centering
\footnotesize
\onehalfspacing
\caption{OLS 1-digit SIC elasticities using ABS, 2008 - 2019.}
\begin{tabular}{ccccccccccc}
\toprule
&(1)&(2)&(3)&(4)&(5)&(6)&(7)&(8)&(9)&(10) \\
\midrule
$\beta_m$         &    0.593\sym{***}&    0.737\sym{***}&    0.756\sym{***}&    0.724\sym{***}&    0.800\sym{***}&    0.815\sym{***}&    0.768\sym{***}&    0.618\sym{***}&    0.639\sym{***}&    0.900\sym{***}\\
          &  (0.168)         &  (0.019)         &  (0.009)         &  (0.021)         &  (0.014)         &  (0.013)         &  (0.032)         &  (0.029)         &  (0.016)         &  (0.038)         \\
[1em]
$\beta_\ell$          &    0.486\sym{**} &    0.107\sym{***}&    0.190\sym{***}&    0.203\sym{***}&    0.034         &    0.101\sym{***}&    0.345\sym{***}&    0.323\sym{***}&    0.312\sym{***}&   -0.065         \\
          &  (0.198)         &  (0.016)         &  (0.008)         &  (0.017)         &  (0.037)         &  (0.009)         &  (0.023)         &  (0.028)         &  (0.016)         &  (0.042)         \\
[1em]
$\beta_k$         &    0.145\sym{**} &    0.131\sym{***}&    0.057\sym{***}&    0.069\sym{***}&    0.142\sym{***}&    0.057\sym{***}&   -0.042\sym{**} &    0.071\sym{***}&   -0.040\sym{***}&    0.076\sym{***}\\
          &  (0.058)         &  (0.016)         &  (0.009)         &  (0.016)         &  (0.027)         &  (0.011)         &  (0.020)         &  (0.018)         &  (0.015)         &  (0.024)         \\
\midrule
LB    &    1.036         &    0.962         &    0.995         &    0.978         &    0.951         &    0.958         &    1.039         &    0.981         &    0.889         &    0.872         \\
UB    &    1.410         &    0.989         &    1.010         &    1.015         &    1.002         &    0.988         &    1.101         &    1.044         &    0.932         &    0.950         \\
RTS       &    1.223         &    0.975         &    1.002         &    0.996         &    0.977         &    0.973         &    1.070         &    1.013         &    0.911         &    0.911         \\
$N\times T$         &       31         &      224         &      511         &      165         &       84         &      187         &      129         &      219         &      280         &       82         \\
$R^2$        &    0.977         &    0.993         &    0.993         &    0.995         &    0.994         &    0.993         &    0.976         &    0.972         &    0.975         &    0.972         \\

\bottomrule
\end{tabular}
\label{tab:ols_abs_sectoral}
\caption*{Standard errors in parentheses. \sym{*} \(p<0.1\), \sym{**} \(p<0.05\), \sym{***} \(p<0.01\). UB and LB are the RTS 95\% upper and lower bounds, obtained using the delta method for combining estimated coefficients.}
\end{table}
}

\begin{figure}[H]
  \centering
  \includegraphics[width=\linewidth]{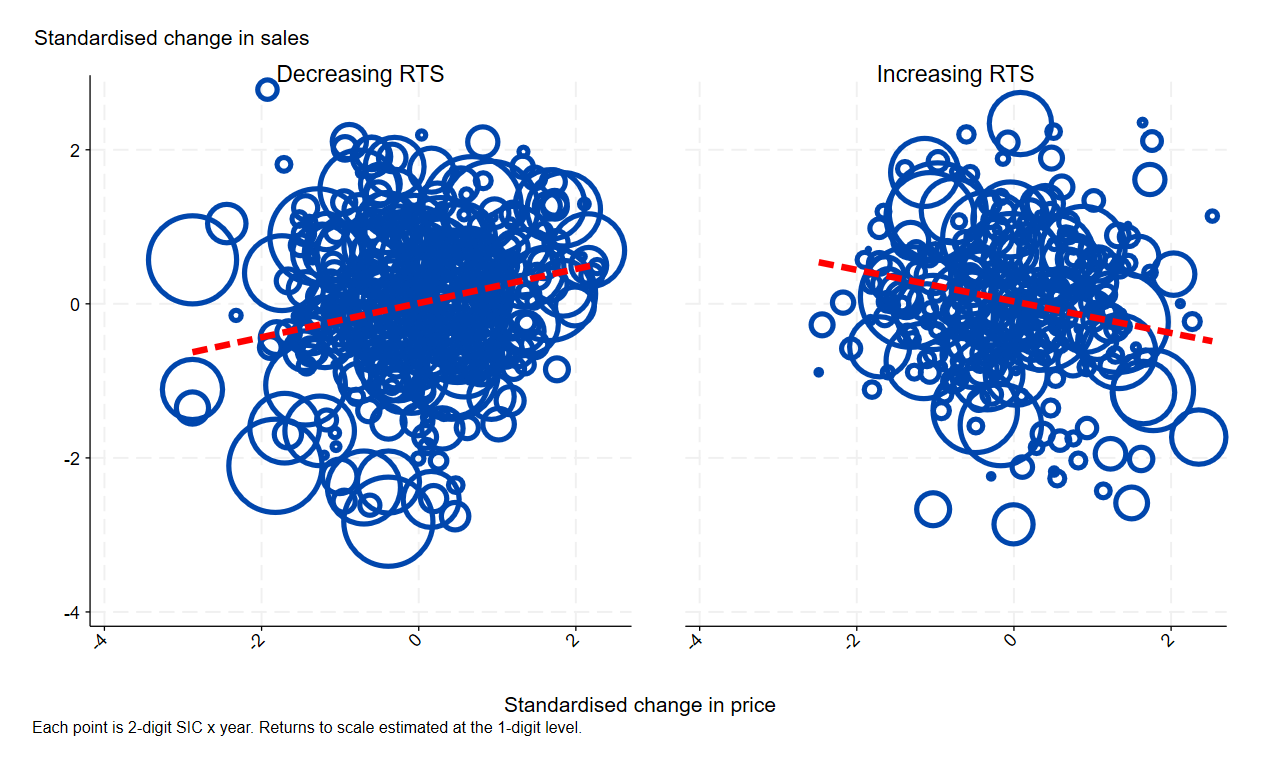}
  \caption{Scatter of 2-digit x year change in industry-level price deflator against standardised change in sales, split by returns to scale (estimated at 1-digit level).}
  \label{fig:standardised_scatter}
\end{figure}

We split the data (at the 2-digit by year level) by the returns to scale of the 1-digit industry they are in. Approximately 45\% of our sector-year observations are in `increasing' returns sectors. Figure~\ref{fig:standardised_scatter} presents the scatter of annual changes in sales against annual changes in output price deflators for those in decreasing or increasing scale sectors. Changes in prices and sales are standardised at the 2-digit SIC level (i.e., we plot sectoral changes, relative to the sector-specific mean, divided by sector-specific standard deviation). We remove the top and bottom 5\% of changes in sales or prices for readability, but the results hold with or without outlier removal. There is a strong positive correlation when there are decreasing RTS, and there is a weak negative relationship for increasing RTS industries that does not hold without standardisation. 
The scatter plot suggest that the relationship between sales and price changes depends on returns to scale. In Table~\ref{tab:delta_price_delta_sales_rts_interaction} we test this hypothesis by regressing differenced deflators on differenced sales, interacting the latter with an indicator for whether industries are in increasing RTS 1-digit sectors. Table~\ref{tab:delta_price_delta_sales_rts_interaction} presents the results for alternative specifications, with year fixed effects, removing outliers (in changes in sales and prices), and weighting by gross value-added or number of firms. 
{
\def\sym#1{\ifmmode^{#1}\else\(^{#1}\)\fi}
\begin{table}[H]
\centering
\onehalfspacing
\caption{Changes in sales against price deflators at 2-digit SIC, ABS 2008 - 2019.}
\begin{tabular}{p{4cm}*{5}{c}}
\toprule
          &\multicolumn{1}{c}{(1)}&\multicolumn{1}{c}{(2)}&\multicolumn{1}{c}{(3)}&\multicolumn{1}{c}{(4)}&\multicolumn{1}{c}{(5)}\\
          &\multicolumn{5}{c}{$\Delta$ prices}\\
\midrule
$\Delta$ sales  &   0.154\sym{***}&    0.160\sym{**} &    0.147\sym{**} &    0.256\sym{***} &   0.269\sym{***} \\
 &  (0.053)    &  (0.052)  &  (0.053) &  (0.064)  & (0.076)    \\
[1em]
$\mathbf{1}\{ RTS > 1 \}  \times \Delta $ sales & -0.142\sym{*}  &   -0.166\sym{**}  &   -0.178\sym{**}  &   -0.294\sym{***} & -0.444\sym{***}  \\
  &  (0.080)  &  (0.072)   &  (0.073)   &  (0.089) & (0.117) \\
\midrule
Year FE & & \checkmark & \checkmark & \checkmark & \checkmark \\
Remove outliers & & & \checkmark & \checkmark & \checkmark \\
GVA Weighted & & & & \checkmark & \\
\#Firms Weighted & & & & & \checkmark \\
\midrule
$N$     &   662  & 662  &   646  & 646  & 646   \\
$R^2$   &  0.014  &    0.222  &  0.213  & 0.280 & 0.277   \\
\bottomrule
\multicolumn{5}{l}{\footnotesize Standard errors in parentheses. \sym{*} \(p<0.1\), \sym{**} \(p<0.05\), \sym{***} \(p<0.01\).} \\
\end{tabular}
\label{tab:delta_price_delta_sales_rts_interaction}
\end{table}
}

There is a strong and significant correlation between changes in sales and prices for sectors with returns to scale below unity. However, this relationship is offset when scale is increasing. The point estimates are consistently negative and large, but statistical significance depends on placing more weight on sectors with a greater share of gross value-added in the economy.

The first row capture the effect with decreaing returns, whilst the sum of the first row and the second row captures the effect with increasing returns. Typically, when there are increasing returns the relationship is close to zero. The cost advantages of scale mean that an increase in output does not lead to price pressures. With decreasing returns, there is typically a strong effect of higher sales raising prices.

\section{Conclusion}

\textcite{ConlonMillerOtgonYao2023_AEApp} pose the question: rising markups, rising prices? But find little evidence in support of this. We ask: rising marginal costs, rising prices? We find evidence in favour of this for decreasing returns to scale industries, but not for increasing returns to scale industries. With decreasing returns, positive demand shocks increase marginal costs and increase prices, hence markups do not necessarily respond, which could account for no relationship between prices and markups. However, with increasing returns, a positive demand shock reduces marginal cost but price is stable, hence markups necessarily increase to sustain price.

The aim of our paper is to provide some initial evidence on the role of costs in determining prices as suggested by \textcite{ConlonMillerOtgonYao2023_AEApp, Syverson2019_JEP}. Further research should expand these initial insights to address the endogeneity issues inherent in reduced-form analyses. A promising direction is to develop a microfounded structural model that would determine each of the key variables (prices, marginal costs, markups) endogenously in response to shocks. This would offer an internally consistent interpretation of the empirical findings in this paper. 

Our analysis implies that discussions of price changes in macroeconomics should consider how underlying market structures affect costs, rather than focusing only on demand-side factors affecting competition through the markup. This is particularly relevant as new artificial intelligence (AI) technologies are emerging, which appear to benefit from scale economies and may affect appropriate monetary policy responses. 

\printbibliography

\appendix

\section{Data}

The ONS releases aggregated production data at different SIC levels via the \href{https://www.ons.gov.uk/businessindustryandtrade/business/businessservices/datasets/uknonfinancialbusinesseconomyannualbusinesssurveysectionsas/current}{Non-financial business economy, UK: Sections A to S} dataset. Our analysis is based on the `ABS standard extracts 2019', which are provided on request by the ONS up to 2019 (further detail at \href{https://www.ons.gov.uk/businessindustryandtrade/business/businessservices/methodologies/annualbusinesssurveyabs}{ABS methodology}). The ABS is the UK's annual production survey which is used to construct national accounts. The ABS surveys a representative sample of firms, roughly 50,000 annually, about various production features such as their sales, input expenditures, and investments. 

At the UK level, the extract covers a wider range of variables than the published data, and it records variables down to the five-digit SIC level. There is a regional extract which covers five main variables down to three-digit group level.

The data that is available from the ONS which we refer to in the paper are here:
\begin{itemize}
    \item Deflators: \url{https://www.ons.gov.uk/economy/inflationandpriceindices/datasets/experimentalindustrydeflatorsuknonseasonallyadjusted}, 
    \item Supply and use tables: \url{https://www.ons.gov.uk/economy/nationalaccounts/supplyandusetables/datasets/supplyanduseofproductsandindustrygvaukexperimental}
    \item Capital stocks: \url{https://www.ons.gov.uk/economy/nationalaccounts/uksectoraccounts/datasets/capitalstocksconsumptionoffixedcapital}.
\end{itemize}

\section{Additional Regression Results}

In this section we provide alternative methods to estimate the production function. We find results that are broadly consistent with our main OLS approach.

Table \ref{tab:ols_abs_sectoral_2} shows estimated output elasticities and returns to scale at the 1-digit SIC level from 3-digit $\times$ year sectoral data, using instrumental variables with lagged inputs as instruments.
{
\def\sym#1{\ifmmode^{#1}\else\(^{#1}\)\fi}
\begin{table}[H]
\centering
\footnotesize
\onehalfspacing
\caption{IV 1-digit SIC elasticities using ABS, 2008 - 2019.}
\begin{tabular}{ccccccccccc}
\toprule
&(1)&(2)&(3)&(4)&(5)&(6)&(7)&(8)&(9)&(10) \\
\midrule
$\beta_m$         &    1.318         &    0.685\sym{***}&    0.731\sym{***}&    0.686\sym{***}&    0.798\sym{***}&    0.779\sym{***}&    0.750\sym{***}&    0.597\sym{***}&    0.663\sym{***}&    0.925\sym{***}\\
          &  (1.246)         &  (0.021)         &  (0.011)         &  (0.027)         &  (0.008)         &  (0.015)         &  (0.031)         &  (0.027)         &  (0.018)         &  (0.059)         \\
[1em]
$\beta_\ell$         &   -0.690         &    0.069\sym{***}&    0.184\sym{***}&    0.202\sym{***}&    0.036\sym{*}  &    0.101\sym{***}&    0.355\sym{***}&    0.340\sym{***}&    0.302\sym{***}&   -0.073         \\
          &  (1.848)         &  (0.016)         &  (0.011)         &  (0.024)         &  (0.021)         &  (0.012)         &  (0.025)         &  (0.024)         &  (0.021)         &  (0.074)         \\
[1em]
$\beta_k$         &    0.687         &    0.222\sym{***}&    0.085\sym{***}&    0.104\sym{***}&    0.166\sym{***}&    0.081\sym{***}&   -0.054\sym{***}&    0.087\sym{***}&   -0.057\sym{***}&    0.059         \\
          &  (0.775)         &  (0.026)         &  (0.014)         &  (0.025)         &  (0.016)         &  (0.013)         &  (0.021)         &  (0.016)         &  (0.020)         &  (0.050)         \\
\midrule
RTS LB    &    0.950         &    0.964         &    0.992         &    0.971         &    0.989         &    0.947         &    1.029         &    0.999         &    0.889         &    0.868         \\
RTS UB    &    1.678         &    0.988         &    1.007         &    1.013         &    1.013         &    0.975         &    1.076         &    1.047         &    0.927         &    0.953         \\
RTS       &    1.314         &    0.976         &    0.999         &    0.992         &    1.001         &    0.961         &    1.052         &    1.023         &    0.908         &    0.911         \\
N $\times$ T         &       19         &      179         &      391         &      125         &       62         &      154         &       94         &      184         &      249         &       73         \\
R$^2$        &    0.957         &    0.994         &    0.994         &    0.995         &    0.998         &    0.992         &    0.990         &    0.978         &    0.973         &    0.971         \\
\bottomrule
\end{tabular}
\label{tab:ols_abs_sectoral_2}
\caption*{Standard errors in parentheses. \sym{*} \(p<0.1\), \sym{**} \(p<0.05\), \sym{***} \(p<0.01\). UB and LB are the RTS 95\% upper and lower bounds, obtained using the delta method for combining estimated coefficients.}
\end{table}
}

Table \ref{tab:ols_abs_sectoral_3} shows estimated output elasticities and returns to scale at the 1-digit SIC level from 3-digit $\times$ year sectoral data, using the control function method of \textcite{LevinsohnPetrin2003_ReStud}.
{
\def\sym#1{\ifmmode^{#1}\else\(^{#1}\)\fi}
\begin{table}[H]
\centering
\footnotesize
\onehalfspacing
\caption{\textcite{LevinsohnPetrin2003_ReStud} 1-digit SIC elasticities using ABS, 2008 - 2019.}
\begin{tabular}{ccccccccccc}
\toprule
&(1)&(2)&(3)&(4)&(5)&(6)&(7)&(8)&(9)&(10) \\
\midrule
$\beta_m$         &    0.637\sym{***}&    0.591\sym{***}&    0.770\sym{***}&    0.654\sym{***}&    0.817\sym{***}&    0.730\sym{***}&    0.788\sym{***}&    0.603\sym{***}&    0.407\sym{***}&    0.745\sym{***}\\
          &  (0.084)         &  (0.086)         &  (0.030)         &  (0.042)         &  (0.043)         &  (0.053)         &  (0.082)         &  (0.061)         &  (0.105)         &  (0.127)         \\
[1em]
$\beta_\ell$         &    0.304         &    0.103\sym{***}&    0.184\sym{***}&    0.240\sym{***}&    0.156\sym{***}&    0.115\sym{***}&    0.331\sym{***}&    0.357\sym{***}&    0.301\sym{***}&   -0.055         \\
          &  (0.207)         &  (0.028)         &  (0.021)         &  (0.045)         &  (0.049)         &  (0.023)         &  (0.068)         &  (0.087)         &  (0.039)         &  (0.136)         \\
[1em]
$\beta_k$         &    0.096         &    0.051         &    0.043\sym{***}&    0.032\sym{*}  &    0.003         &    0.049\sym{*}  &    0.061         &    0.105\sym{**} &   -0.014         &    0.040         \\
          &  (0.070)         &  (0.042)         &  (0.010)         &  (0.016)         &  (0.074)         &  (0.025)         &  (0.047)         &  (0.044)         &  (0.029)         &  (0.030)         \\

\midrule
RTS LB    &    0.503         &    0.523         &    0.935         &    0.798         &    0.819         &    0.780         &    1.008         &    0.878         &    0.453         &    0.444         \\
RTS UB    &    1.570         &    0.968         &    1.059         &    1.051         &    1.132         &    1.007         &    1.351         &    1.253         &    0.934         &    1.016         \\
RTS       &    1.037         &    0.745         &    0.997         &    0.925         &    0.976         &    0.893         &    1.180         &    1.065         &    0.694         &    0.730         \\
N $\times$ T         &       31         &      222         &      495         &      161         &       73         &      189         &      119         &      220         &      278         &       82         \\
\bottomrule
\end{tabular}
\label{tab:ols_abs_sectoral_3}
\caption*{Standard errors in parentheses. \sym{*} \(p<0.1\), \sym{**} \(p<0.05\), \sym{***} \(p<0.01\). UB and LB are the RTS 95\% upper and lower bounds, obtained using the delta method for combining estimated coefficients.}
\end{table}
}

\end{document}